\documentclass[preprintnumbers]{revtex4}

\usepackage{graphicx}
\usepackage{epsf}
\usepackage{amsmath,amssymb}
\newcommand{\bvec}{\boldsymbol}

\begin{document}
\title{Isovector and isoscalar dipole excitations in $^{9}$Be and $^{10}$Be studied with
antisymmetrized molecular dynamics}
\author{Yoshiko Kanada-En'yo}
\affiliation{Department of Physics, Kyoto University, Kyoto 606-8502, Japan}
\begin{abstract}
Isovector and isoscalar dipole excitations in $^9$Be and $^{10}$Be
are investigated in the framework of antisymmetrized molecular dynamics, 
in which angular-momentum and 
parity projections are performed. 
In the present method, 1p-1h excitations on the ground state and 
large amplitude $\alpha$-cluster mode are incorporated.  
The isovector giant dipole resonance (GDR) in $E>20$ MeV 
shows the two peak structure which is understood by the dipole excitation 
in the 2$\alpha$ core part with the prolate deformation.
Because of valence neutron modes against the $2\alpha$ core, 
low-energy E1 resonances appear in $E<20$ MeV 
exhausting about $20\%$ of the Thomas-Reiche-Kuhn sum rule and $10\%$ of the calculated 
energy-weighted sum. 
The dipole resonance at $E\sim 15$ MeV in $^{10}$Be can be interpreted 
as the parity partner of the ground state having a $^6$He+$\alpha$ structure
and has the remarkable E1 strength 
because of coherent contribution of two valence neutrons.
The ISD strength for some low-energy resonances are significantly enhanced
by the coupling with the $\alpha$-cluster mode. 
The calculated E1 strength of $^9$Be reasonably describes the global feature of
experimental photonuclear cross sections consisting of    
the low-energy strength in $E<20$ MeV and the GDR in $E>20$ MeV.
\end{abstract}
\maketitle
\section{Introduction}
In neutron-rich nuclei, various 
exotic phenomena appear because of excess neutrons. 
One of the current issues
concerning exotic excitation modes in neutron-rich nuclei is low-energy dipole excitations
 \cite{kobayashi89,Ieki:1992mc,Sackett:1993zz,Shimoura:1994me,Zinser:1997da,Aumann:1999mb,Nakamura:2006zz,
Kanungo:2015dna,Millener:1983zz,
Hansen:1987mc,Bertsch:1990zza,Suzuki:1990uq,Honma90,Bertsch:1991zz,Sagawa92,Csoto:1994ji,Suzuki00,Garrido:2002ws,Myo:2003bh,
chulkov,Bertulani:2007rm,Hagino:2007rn,Hagino:2009sj,Baye:2009zz,Kikuchi:2010zzb,Pinilla:2012zz,Kikuchi:2013ula,Nakamura,Palit,
Fukuda:2004ct,Myo98,Sagawa:2001kx,
Tohyama95,Hamamoto96a,Hamamoto96b,Sagawa:1999zz,Colo:2001fz,Nakatsukasa:2004ys,KanadaEn'yo:2005wd,
VanIsacker:1992zz,Catara,Matsuo:2001wy,Vretenar:2001hs,Goriely:2002cx,Paar:2002gz,Tsoneva:2003gv,Matsuo:2004pr,
Piekarewicz:2006ip,Terasaki:2006ts,Liang:2007ri,Tsoneva:2007fk,Paar:2007bk,Yoshida:2008rw,Co':2009gi,Martini:2011gy,Inakura:2011mv,
RocaMaza:2011ug,Ebata:2014aaa,Bacca:2014rta,Piekarewicz:2010fa,Carbone:2010az,Reinhard:2012vw,Inakura:2013waa,
Govaert:1998zz,Herzberg:1999tb,Leistenschneider:2001zz,Ryezayeva:2002zz,Tryggestad:2003gz,Hartmann:2004zz,
Adrich:2005zz,Gibelin:2008zz,Klimkiewicz:2007zz,Schwengner:2008rk,Wieland:2009zz,Endres:2009zz,Endres:2010zw,
Tamii:2011pv}.
For stable nuclei, isovector giant dipole resonances (GDR) 
have been systematically observed in various nuclei 
by measurements of photonuclear cross sections
(for example, Ref.~\cite{Berman:1975tt} and references therein). The GDR is understood as
opposite oscillation between protons and neutrons and microscopically 
described by coherent 1p-1h excitations. Peak structure of strength function of the GDR 
has been often discussed in relation with nuclear deformation.
In neutron-rich nuclei, low-energy dipole resonances have been suggested to appear
because of excess neutron motion against a core. 
The so-called soft dipole resonance, which is 
the enhanced E1 strength observed in the extremely low-energy 
($E\le 2-3$ MeV) region in neutron halo nuclei such as $^6$He and $^{11}$Li, has been 
intensively studied by experimental and theoretical groups and often discussed in relation with
two-neutron correlation 
\cite{kobayashi89,Ieki:1992mc,Sackett:1993zz,Shimoura:1994me,Zinser:1997da,Aumann:1999mb,Nakamura:2006zz,Kanungo:2015dna,
Hansen:1987mc,Bertsch:1990zza,Suzuki:1990uq,Honma90,Bertsch:1991zz,Sagawa92,Csoto:1994ji,Suzuki00,Garrido:2002ws,Myo:2003bh,
chulkov,Bertulani:2007rm,Hagino:2007rn,Hagino:2009sj,Baye:2009zz,Kikuchi:2010zzb,Pinilla:2012zz,Kikuchi:2013ula}.
More generally 
the low-energy E1 strength typically in $5\le E \le 15$ MeV region 
is predicted in various nuclei in a
wide mass region, and the one decoupling from the GDR is called pigmy dipole resonance.
The role of excess neutrons in and the collectivity of these low-energy dipole excitations 
are topics of interest in theoretical studies.
The low-energy E1 strength has been also discussed in relation with neutron skin thickness 
and density dependence of the symmetry energy, though the correlation between 
pigmy dipole strength, neutron skin thickness, and the symmetry energy is under debate
\cite{Tsoneva:2003gv,Piekarewicz:2006ip,
Piekarewicz:2010fa,Carbone:2010az,Reinhard:2012vw,Inakura:2013waa,Klimkiewicz:2007zz,Tamii:2011pv}.

The experimental measurements of the low-energy E1 strength in $^9$Be 
by photodisintegration have been performed mainly in astrophysical interests
\cite{Jakobson:1961zz,hughes75,Goryachev92,Utsunomiya:2001sz,Arnold:2011nv}.
In these years, precise data of the E1 strength for low-lying positive-parity states of $^9$Be 
have been reported \cite{Arnold:2011nv}.
Moreover, the recently measured photodisintegration cross sections of $^9$Be indicate 
the significant low-energy E1 strength around $E=10$ MeV exhausting 
about 10\% of the Thomas-Reiche-Kuhn (TRK)  sum rule \cite{Utsunomiya15}, 
consistently with the bremsstrahlung data \cite{Goryachev92}.
The experimental data of the E1 strength of $^9$Be are available in a wide energy region 
from low energy to high energy:
the E1 strength for the positive-parity states in $E\le 5$ MeV \cite{Arnold:2011nv}, 
the significant E1 strength around $E=10$ MeV \cite{Utsunomiya15,Goryachev92}, and 
the E1 strength in $E>20$ MeV for the GDR \cite{Ahrens:1975rq}.

Our aim is to investigate isovector and isoscalar dipole strengths in $^9$Be and $^{10}$Be to 
understand the low-energy dipole modes. I try to clarify 
the role of excess neutrons and the decoupling mechanism of the low-energy dipole modes 
from the GDR.
The low-lying states of $^9$Be are understood by $2\alpha+n$ cluster structures
as discussed in cluster models \cite{Okabe77,Okabe78,Fonseca79,Descouvemont:1989zz,Arai:1996dq}. 
The photodisintegration cross sections in very low-energy region of $^9$Be
have been theoretically investigated by
two-body ($^8$Be+$n$) and three-body ($2\alpha+n$) cluster 
models with continuum states
\cite{Efros:1998yk,Arai:2003jm,Burda:2010an,Garrido:2010xz,
AlvarezRodriguez:2010ng,Efros:2013zoa,Odsuren:2015pja}.
However, such cluster models are not able to describe
the high-energy dipole strength of the GDR, to which 
coherent 1p-1h excitations may contribute.
To investigate the pigmy and giant dipole resonances in general nuclei,  
shell model and mean-field approaches have been applied.
The former may not be suitable for such largely deformed nuclei as
Be isotopes having cluster structures.
The latter is usually based on the random phase approximation (RPA) with and without continuum. 
Although the RPA calculation is 
successful for a variety of collective excitations in heavy mass nuclei, 
it is a small amplitude approximation neglecting large amplitude motion. 
Moreover, in most of current mean-field approaches, 
the RPA calculation is based a parity-symmetric mean field in a strong coupling picture
without the angular-momentum and parity projections, and the 
coupling of single-particle excitations in the mean-field with rotation and parity transformation is 
not taken into account microscopically.

To take into account the coherent 1p-1h excitations and the   
large amplitude cluster mode as well as the angular-momentum and parity projections, we develop a method of antisymmetrized 
molecular dynamics (AMD) \cite{Ono:1991uz,Ono:1992uy,KanadaEnyo:1995tb,
KanadaEnyo:1995ir,AMDsupp,KanadaEn'yo:2012bj}. 
The time-dependent AMD, which was originally developed 
for study of heavy-ion reactions \cite{Ono:1991uz,Ono:1992uy}, was applied to 
investigate E1 and monopole excitations \cite{KanadaEn'yo:2005wd,Furuta:2010ad}. 
However, in the time-dependent AMD approach,
the angular-momentum and parity projections are not performed, and therefore,  
ground state structures and the coupling of single-particle excitations 
with the rotational motion are not sufficiently described. 
Instead of the time-dependent AMD, 
we superpose the angular-momentum and parity projected wave functions of 
various configurations including the 1p-1h and cluster excitations. 
We first perform the variation after the angular-momentum and parity projections in the AMD 
framework (AMD+VAP) \cite{KanadaEn'yo:1998rf,KanadaEn'yo:1999ub,KanadaEn'yo:2006ze}
to obtain the ground state wave function. 
Then we describe small amplitude motions by taking into 
account 1p-1h excitations on the obtained ground 
state wave function with the shifted AMD method
as done  in Ref.~\cite{Kanada-En'yo:2013dma} for monopole excitations of $^{16}$O.
To incorporate the large amplitude cluster motion, we combine the 
the generator coordinate method (GCM) with the shifted AMD
by superposing $^{5,6}\textrm{He}+\alpha$ cluster wave functions. 
The angular-momentum and parity projections are performed in the present framework.
Applying the present method, I investigate the dipole transitions
$^8\textrm{Be}(0^+_1)\to ^8\textrm{Be}(1^-)$, 
$^9{\rm Be}(3/2^-_1)\to ^9{\rm Be}(1/2^+, 3/2^+, 5/2^+)$, and 
$^{10}\textrm{Be}(0^+_1)\to ^{10}\textrm{Be}(1^-)$.

This paper is organized as follows. 
The present method of AMD is formulated in section \ref{sec:formulation}, 
and section \ref{sec:results} discusses the ground state structures and the E1 and isoscalar 
dipole (ISD) excitations in Be isotopes. 
The paper concludes with a summary in section \ref{sec:summary}.

\section{Formulation of AMD for dipole excitations}
\label{sec:formulation}

I apply the AMD+VAP method 
to obtain $A$-nucleon wave functions for the ground states of $^8$Be, $^9$Be, and
$^{10}$Be.
To investigate dipole excitations, I incorporate
1p-1h excitations on the ground state wave function 
with the shifted AMD method. 
I also perform the $\alpha$-cluster GCM calculation combined with the shifted AMD
to see how the $\alpha$-cluster mode affects the dipole excitations. 
In this section, I explain the formulation of the AMD+VAP, the 
shifted AMD, and the $\alpha$-cluster GCM calculations, and also describe the definition of the 
dipole strengths.

 \subsection{AMD wave function}
An AMD wave function is given by a Slater determinant,
\begin{equation}
 \Phi_{\rm AMD}({\bvec{Z}}) = \frac{1}{\sqrt{A!}} {\cal{A}} \{
  \varphi_1,\varphi_2,...,\varphi_A \},\label{eq:slater}
\end{equation}
where  ${\cal{A}}$ is the antisymmetrizer. The $i$th single-particle wave function 
$\varphi_i$ is written by a product of
spatial, spin, and isospin
wave functions as
\begin{eqnarray}
 \varphi_i&=& \phi_{{\bvec{X}}_i}\chi_i\tau_i,\\
 \phi_{{\bvec{X}}_i}({\bvec{r}}_j) & = &  \left(\frac{2\nu}{\pi}\right)^{4/3}
\exp\bigl\{-\nu({\bvec{r}}_j-\bvec{X}_i)^2\bigr\},
\label{eq:spatial}\\
 \chi_i &=& (\frac{1}{2}+\xi_i)\chi_{\uparrow}
 + (\frac{1}{2}-\xi_i)\chi_{\downarrow}.
\end{eqnarray}
$\phi_{{\bvec{X}}_i}$ and $\chi_i$ are the spatial and spin functions, respectively, and 
$\tau_i$ is the isospin
function fixed to be up (proton) or down (neutron). 
The width parameter $\nu$ is fixed to be the optimized value 
for each nucleus.
To separate the center of mass motion from the total wave function $\Phi_{\rm AMD}({\bvec{Z}})$,
the following condition should be satisfied,  
\begin{equation}\label{eq:cm}
\frac{1}{A}\sum_{i=1,\ldots,A} \bvec{X}_i=0.
\end{equation}
In the present calculation, I keep this condition and exactly remove the contribution of
the center of mass motion.

Accordingly, an AMD wave function
is expressed by a set of variational parameters, ${\bvec{Z}}\equiv 
\{{\bvec{X}}_1,\ldots, {\bvec{X}}_A,\xi_1,\ldots,\xi_A \}$,
which specify centroids of single-nucleon Gaussian wave packets and spin orientations 
for all nucleons. In the AMD framework, existence of clusters is not assumed {\it a priori} 
because Gaussian centroids, ${\bvec{X}}_1,\ldots,{\bvec{X}}_A$ of
all single-nucleon wave packets are independently 
treated as variational parameters. Nevertheless, 
a multi-center cluster wave function can be described by the 
AMD wave function with the corresponding configuration of Gaussian centroids. 
It should be commented that the AMD wave function is similar to the wave function used in Fermionic molecular dynamics calculations \cite{Feldmeier:1994he,Neff:2002nu}.

\subsection{AMD+VAP}
In the AMD+VAP method, the parameters ${\bvec{Z}}= 
\{{\bvec{X}}_1,{\bvec{X}}_2,\ldots, {\bvec{X}}_A,\xi_1,\xi_2,\ldots,\xi_A \}$
in the AMD wave function
are determined by the energy variation 
after the angular-momentum and pariy projections (VAP).
It means that, 
${\bvec{X}}_i$ and $\xi_{i}$ for the lowest $J^\pi$ state 
are determined so as to minimize the energy expectation value of the Hamiltonian
for the $J^\pi$-projected AMD wave function; 
\begin{eqnarray}
&& \frac{\delta}{\delta{\bvec{X}}_i}
\frac{\langle \Phi|H|\Phi\rangle}{\langle \Phi|\Phi\rangle}=0,\\
&& \frac{\delta}{\delta\xi_i}
\frac{\langle \Phi|H|\Phi\rangle}{\langle \Phi|\Phi\rangle}=0,\\
&&\Phi= P^{J\pi}_{MK}\Phi_{\rm AMD}({\bvec{Z}}),
\end{eqnarray}
where $P^{J\pi}_{MK}$ is the angular-momentum and parity projection operator. 
After the VAP calculation,  
the optimized parameters $\bvec{Z}^{J\pi}_\textrm{VAP}$ for the lowest $J^\pi$ state
are obtained.
For the ground state, the VAP with $J^\pi=0^+$ and $K=0$ is performed
for $^8$Be and $^{10}$Be, and that with 
$J^\pi=3/2^-$ and  $K=3/2$ is done for $^9$Be.
I rewrite the parameters $\bvec{Z}^{J\pi}_\textrm{VAP}$ obtained by the VAP 
for the ground state as $\bvec{Z}^0=\{\bvec{X}^0_1,\ldots,\xi^0_{1},\ldots\}$.

\subsection{Shifted AMD}

To taken into account 1p-1h excitations, 
I consider small variation of single-particle wave functions in the 
ground state wave function $\Phi_{\rm AMD}({\bvec{Z}^0})$
by shifting the position of the Gaussian centroid 
of the $i$th single-particle wave function,
${\bvec{X}}^0_i\rightarrow {\bvec{X}}^0_i+\epsilon{\bvec{e}}_\sigma$, where
$\epsilon$ is a small constant and 
${\bvec{e}}_\sigma$ is an unit vector with the label $\sigma$. In the present calculation, 
${\bvec{e}}_1,\ldots,{\bvec{e}}_8$ for 8 directions are adopted to obtain the approximately converged result for the E1 and ISD strengths.
Details of the adopted unit vectors ${\bvec{e}}_\sigma$ ($\sigma=1,\ldots,8$) are described 
in section \ref{sec:results}.
For the spin part, 
I consider the spin-nonflip single-particle state $\chi_i$
and the spin-flip state $\bar\chi_i$ ($\langle\bar\chi_i|\chi_i\rangle=0$),
\begin{equation}
\bar\chi_i = (\frac{1}{2}+\bar\xi_i)\chi_{\uparrow}
 + (\frac{1}{2}-\bar\xi_i)\chi_{\downarrow},
\end{equation} 
where $\bar\xi_i=-1/4\xi^*_i$.
For all single-particle wave functions, I consider spin-nonflip and spin-flip
states shifted to eight directions independently and prepare 
$16A$ AMD wave functions, 
$\Phi_{\rm AMD}({\bvec{Z}}_{\rm nonflip}^0(i,\sigma))$
and $\Phi_{\rm AMD}({\bvec{Z}}_{\rm flip}^0(i,\sigma))$, with the shifted parameters
\begin{eqnarray}
&&\bvec{Z}_{\rm nonflip}^0(i,\sigma)\equiv 
\{{\bvec{X}^0_1}',\cdots,{\bvec{X}^0_i}'+\epsilon {\bvec{e}}_\sigma,\cdots,
{\bvec{X}^0_A}',\xi^0_1,\cdots,\xi^0_i,\cdots,\xi^0_A \},\\
&&\bvec{Z}_{\rm flip}^0(i,\sigma)\equiv 
\{{\bvec{X}^0_1}',\cdots,{\bvec{X}^0_i}'+\epsilon {\bvec{e}}_\sigma,\cdots,
{\bvec{X}^0_A}',\xi^0_1,\cdots,\bar\xi^0_i,\cdots,\xi^0_A \}.
\end{eqnarray}
Here ${{\bvec{X}}^{0}_j}'$ is chosen to be 
${{\bvec{X}}^{0}_j}'={\bvec{X}}^{0}_j-\epsilon {\bvec{e}}_\sigma/(A-1)$ to take into account the recoil effect so that 
the center of mass motion is separated exactly.
Those shifted AMD wave functions 
$\Phi_{\rm AMD}({\bf Z}_{\rm (non)flip}^0(i,\sigma))$ 
and the original wave function
$\Phi_{\rm AMD}({\bf Z}^0)$
are superposed to obtain the final wave functions for the ground and excited states, 
\begin{eqnarray}
\Psi^{\rm sAMD}_{{\rm Be}(J^\pi_k)}&=&\sum_K c_0(J^\pi_k; K) 
P^{J\pi}_{MK}\Phi_{\rm AMD}({\bf Z}^0)\\
&&+\sum_{i=1,\ldots,A}\sum_{\sigma}\sum_K 
c_1(J^\pi_k; i,\sigma,K) P^{J\pi}_{MK}\Phi_{\rm AMD}({\bf Z}_{\rm nonflip}^0(i,\sigma))\\
&&+
\sum_{i=1,\ldots,A}\sum_{\sigma}\sum_K 
c_2(J^\pi_k; i,\sigma,K) P^{J\pi}_{MK}\Phi_{\rm AMD}({\bf Z}_{\rm flip}^0(i,\sigma)),
\end{eqnarray}
where the coefficients $c_0$, $c_1$, and $c_2$ are determined by diagonalization 
of the norm and Hamiltonian matrices.
I call this method ``the shifted AMD'' (sAMD). 
The spin-nonflip version of the sAMD has been applied to
investigate monopole excitations of $^{16}$O in Ref.~\cite{Kanada-En'yo:2013dma}.

I choose an enough small value of
the spatial shift $\epsilon$, typically $\epsilon=0.1$ fm, so as 
to obtain $\epsilon$-independent results.
The model space of the sAMD contains
the 1p-1h excitations that
are written by a small shift of a single-nucleon Gaussian wave function
of the ground state wave function. In the intrinsic frame 
before the angular-momentum and parity projections, the ground state AMD wave function is expressed by 
a Slater determinant, and therefore,  
the sAMD method corresponds to the RPA
in the restricted model space of the linear combination of shifted Gaussian wave functions. 
However, since the projected states are 
superposed in the sAMD, the coupling of the
1p-1h excitations with the rotation and parity transformation
is properly taken into account. Therefore, the sAMD contains, in principle,
higher correlations beyond the RPA in mean-field approximation.

Note that, the shift $\bvec{X}_i\to \bvec{X}_i+\epsilon\bvec{e}_\sigma$ of the Gaussian wave packet 
can be expressed by a linear combination of harmonic oscillator (h.o.) 
orbits at $\bvec{X}_i$. For instance, when three vectors $\bvec{e}_x$, $\bvec{e}_y$, and $\bvec{e}_z$ are chosen for $\bvec{e}_\sigma$, the linear combination of 
$\phi_{{\bvec{X}}_i}$, $\phi_{{\bvec{X}}_i+\epsilon{\bvec{e}}_x}$, $\phi_{{\bvec{X}}_i+\epsilon{\bvec{e}}_y}$, 
and $\phi_{{\bvec{X}}_i+\epsilon{\bvec{e}}_z}$  in the small $\epsilon$ limit is equivalent to that of 
the $0s$ and $0p$ orbits around $\bvec{X}_i$. It means that, 
in the case that the recoil effect is omitted,
the sAMD can be regarded as an extended AMD method, in which higher h.o. orbits are incorporated
in addition to the default $0s$ orbit at $\bvec{X}_i$
for the $i$th single-particle wave function. In the particular case of $\sigma=x,y,z$, 
it can be called ``$p$-wave AMD''.

\subsection{$\alpha$-cluster GCM}
In the ground state wave functions 
obtained by the AMD+VAP for $^8$Be, $^9$Be, and $^{10}$Be, 
an $\alpha$ cluster is formed even though 
any clusters are not {\it a priori} assumed in the framework. Consequently, 
Gaussian centroids $\bvec{X}^0_i$ for two protons and two neutrons 
are located at almost the same.
The inter-cluster motion of 
$^5$He+$\alpha$ and $^6$He+$\alpha$ structures in $^9$Be and $^{10}$Be
can be excited by the dipole operators. 
To incorporate the large amplitude $\alpha$-cluster mode, we perform the 
$\alpha$-cluster  GCM ($\alpha$GCM) calculation with respect to the inter-cluster distance.
For simplicity, we label four nucleons composing the $\alpha$ cluster 
as $i=1,\ldots,4$ and other nucleons as $i=5,\ldots,A$. 
The center of mass position of the $\alpha$ cluster is localized around
$\bvec{R}_{\alpha}=\frac{1}{4}\textrm{Re}[\bvec{X}^0_1+\bvec{X}^0_2+\bvec{X}^0_3+\bvec{X}^0_4]$. 
The inter-cluster distance $D_\alpha$ is written  as
\begin{equation}
D_\alpha\equiv \left|\textrm{Re}\left[\frac{1}{4}\sum_{i=1,\ldots,4} \bvec{X}^0_i 
-\frac{1}{A-4}\sum_{i=1,\ldots,4} \bvec{X}^0_i \right]\right|=\frac{A}{A-4}R_{\alpha}
\end{equation}
with $R_{\alpha}\equiv|\bvec{R}_{\alpha}|$.
To perform the $\alpha$GCM calculation based on the 
ground state wave function $\Phi_{\rm AMD}(\bvec{Z}^0)$, 
I change the inter-cluster distance $D_\alpha\to D_\alpha+\Delta D$ by shifting positions of single-nucleon Gaussian
centroids
$\bvec{X}_i^0 \to \bvec{X}^0_{i,D_\alpha}(\Delta D)$
by hand as 
\begin{eqnarray}
&&\bvec{X}^0_{i,D_\alpha}(\Delta D)=\bvec{X}^0_i+\frac{A-4}{A}\Delta D\hat{\bvec{R}}_{\alpha} \quad (i\le 4),\\
&&\bvec{X}^0_{i,D_\alpha}(\Delta D)=\bvec{X}^0_i-\frac{4}{A}\Delta D\hat{\bvec{R}}_{\alpha} \quad (i > 4),
\end{eqnarray}
and superpose the wave functions with different $\Delta D$ values. I combine the $\alpha$GCM with the
sAMD and express the total wave function as
\begin{eqnarray}
\Psi^{\textrm{sAMD+}\alpha\textrm{GCM}}_{{\rm Be}(J^\pi_k)}&&=
\sum_K c_0(J^\pi_k; K) 
P^{J\pi}_{MK}\Phi_{\rm AMD}({\bf Z}^0)\nonumber\\
&&+\sum_{i=1,\ldots,A}\sum_{\sigma}\sum_K 
c_1(J^\pi_k; i,\sigma,K) P^{J\pi}_{MK}\Phi_{\rm AMD}({\bf Z}_{\rm nonflip}^0(i,\sigma))\nonumber\\
&&+
\sum_{i=1,\ldots,A}\sum_{\sigma}\sum_K 
c_2(J^\pi_k; i,\sigma,K) P^{J\pi}_{MK}\Phi_{\rm AMD}({\bf Z}_{\rm flip}^0(i,\sigma))\nonumber  \\
&&+\sum_{\Delta D}\sum_K c_3(J^\pi_k; \Delta D,K) 
P^{J\pi}_{MK}\Phi_{\rm AMD}(\bvec{Z}^0_{D_\alpha}(\Delta D))\label{eq:sAMD+aGCM},
\end{eqnarray}
where 
$\bvec{Z}^0_{D_\alpha}(\Delta D)\equiv \left\{\bvec{X}^0_{1,D_\alpha}(\Delta D),
\ldots,\bvec{X}^0_{A,D_\alpha}(\Delta D), \xi_1,\ldots,\xi_A \right\}$. The coefficients are determined by 
diagonalization 
of the norm and Hamiltonian matrices.

\section{Isovector and Isoscalar dipole transitions}

The E1 operator ${\cal M}(E1;\mu)$ is given by the 
isovector dipole operator as
\begin{eqnarray}
{\cal M}(E1;\mu)&=&\frac{N}{A}\sum^\textrm{proton}_i  
r_i Y^1_\mu(\hat{\bvec{r}}_i)-\frac{Z}{A}\sum^\textrm{proton}_i  
r_i Y^1_\mu(\hat{\bvec{r}}_i).
\end{eqnarray}
The ISD operator ${\cal M}(IS1;\mu)$ is 
defined as
\begin{eqnarray}
{\cal M}(IS1;\mu)&=&\sum_i  r^3_i Y^1_\mu(\hat{\bvec{r}}_i),
\end{eqnarray}
which excites the compressive dipole mode. 
The E1 and ISD strengths for the transition $\textrm{g.s.}\to J_k$
are given by the matrix elements of the dipole operators as
\begin{eqnarray}
B(E1;\textrm{g.s.}\to J_k)&=&\frac{1}{2I_\textrm{g.s.}+1}
|\langle J_k||{\cal M}(E1)|| {\rm g.s.}\rangle |^2,\\
B(IS1;\textrm{g.s.}\to J_k)&=&\frac{1}{2I_\textrm{g.s.}+1}
|\langle J_k||{\cal M}(IS1)|| {\rm g.s.}\rangle |^2,
\end{eqnarray}
where $I_\textrm{g.s.}$ is the ground state angular momentum.
The energy-weighted sum (EWS) of the E1 and ISD strengths is defined as
\begin{eqnarray}
S(E1)\equiv \sum_{J_k} E_{J_k} B(E1; \textrm{g.s.}\to J_k),\\
S(IS1)\equiv \sum_{J_k} E_{J_k} B(IS1; \textrm{g.s.}\to J_k),
\end{eqnarray}
where $E_{J_k}$ is the energy of the $J_k$ state.
If the interaction commutes with the E1 operator, $S(E1)$ is identical to 
the TRK sum rule:
\begin{eqnarray}
S({\rm TRK})\equiv \frac{9\hbar^2}{8\pi m}\frac{NZ}{A}.
\end{eqnarray}
Since the nuclear interaction does not commute with the E1 operator,
$S(E1)$ is usually enhanced from $S({\rm TRK})$. 

In the present framework, all excited states are discrete states without escaping widths
because the out-going condition in the asymptotic region
is not taken into account.
I calculate the E1 and ISD strengths for discrete states and smear the
strengths with a Gaussian by hand to obtain dipole strength functions as
\begin{eqnarray}
\frac{dB(E1)}{dE}&=& \sum_J \sum_{k}
\frac{\sqrt{\pi}}{\gamma} \textrm{e}^{-\frac{(E-E_{J_k})^2}{\gamma^2}}
B(E1; \textrm{g.s.}\to J_k),\\
\frac{dB(IS1)}{dE}&=&\sum_J \sum_{k}
\frac{\sqrt{\pi}}{\gamma} \textrm{e}^{-\frac{(E-E_{J_k})^2}{\gamma^2}}
B(IS1; \textrm{g.s.}\to J_k),
\end{eqnarray}
where $\gamma$ is the smearing width. 
The photonuclear  cross section is dominated by E1 transitions and 
related to the E1 strength function as
\begin{equation}
\sigma(E)= \frac{16\pi^3}{9}\frac{e^2}{\hbar c}E\frac{dB(E1)}{dE}.
\end{equation}


\section{Results}\label{sec:results}

\subsection{Effective nuclear interactions}
I use an effective nuclear interaction consisting of the central force of the MV1 force\cite{TOHSAKI} and the
spin-orbit force of the G3RS force \cite{LS1,LS2}, and the Coulomb force.
The MV1 force is given by a two-range Gaussian two-body term and a zero-range three-body term.
The G3RS spin-orbit force is a two-range Gaussian force.
The Bartlett, Heisenberg and Majorana parameters for the case 3 of the MV1 force are  
$b=h=0$ and $m=0.62$, and the strengths of the G3RS spin-orbit force are 
$u_{I}=-u_{II}\equiv u_{ls}=3000$ MeV. These interaction parameters are the same
as those used in 
Refs.~\cite{KanadaEn'yo:1998rf,KanadaEn'yo:1999ub,KanadaEn'yo:2006ze}, 
in which the AMD+VAP calculation describes well 
properties of the ground and excited states 
of $^{10}$Be and $^{12}$C.

\subsection{Ground states}
\begin{figure}[htb]
\begin{center}
\includegraphics[width=7.0cm]{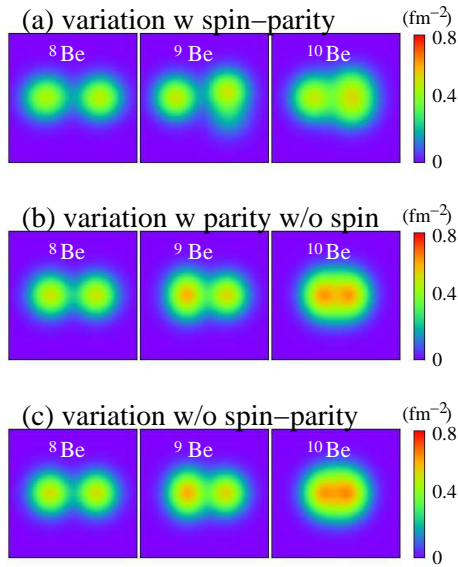} 	
\end{center}
  \caption{(color online)
Density distribution of the intrinsic wave functions of the ground states of $^8$Be, $^9$Be, and $^{10}$Be
obtained by 
(a) the AMD+VAP (variation 
after the angular-momentum and parity projections),  (b) the variation  
after the parity projection without the angular-momentum projection,
and (c) the variation without the  angular-momentum and parity projections.
\label{fig:beiso-dense}}
\end{figure}

I perform the AMD+VAP calculation
to obtain the ground state wave functions  
for $^8$Be($0^+_1$), $^9$Be($3/2^-_1$), and $^{10}$Be($0^+_1$).
The width parameter is chosen to be  
$\nu=0.20$ fm$^{-2}$ for $^8$Be and $^9$Be and  
$\nu=0.19$ fm$^{-2}$ for $^{10}$Be to minimize the ground state energy. 
Figure \ref{fig:beiso-dense}(a) shows intrinsic density distribution of the obtained wave functions
$\Phi_{\rm AMD}({\bvec{Z}^0})$
for the ground states.   As seen in the density,
the $\alpha+\alpha$,  $^5{\rm He}+\alpha$,  and $^6{\rm He}+\alpha$ cluster structures are developed 
in $^8$Be, $^9$Be, and $^{10}$Be, respectively.  
Considering that the $^5{\rm He}$ and  $^6{\rm He}$ clusters have 
$\alpha+n$ and $\alpha+2n$ structures, the ground states of $^9$Be and $^{10}$Be are regarded as 
the $2\alpha$ cluster core with valence neutrons, $2\alpha+n$ and $2\alpha+2n$, in which the valence neutrons
are localized around one of the 2$\alpha$.

As given in Eq.~\eqref{eq:slater}, 
an AMD wave function is expressed by a
single Slater determinant. However, the projected state $P^{J\pi}_{MK}\Phi_{\rm AMD}(\bvec{Z}^0)$ 
for the ground state wave function
contains higher correlations beyond mean-field approximations, 
which is efficiently incorporated 
by the VAP calculation.
Indeed, cluster structures are remarkable in the present VAP result
but they are relatively suppressed in calculations without the
projections. 
For comparison with the present result obtained 
by the VAP (variation after the angular-momentum and parity projections),  
the result obtained by the variation without the angular-momentum and parity projections and 
that after the parity projection without the angular-momentum projection are also demonstrated
in Fig.~\ref{fig:beiso-dense}(b) and (c), 
respectively.
It is clearly seen that  the result of $^{10}$Be obtained by the variation 
without the projections shows weak clustering with a parity-symmetric intrinsic structure
(see the right panels of Fig.~\ref{fig:beiso-dense}(b) and (c)).
This indicates that the angular-momentum and parity projections in the energy variation is 
essential to obtain the parity-asymmetric structure with the $^6$He+$\alpha$ correlation in $^{10}$Be. 

The root mean square radii of point-proton distribution
of $^8$Be, $^9$Be, and $^{10}$Be calculated by the AMD+VAP 
are 2.73 fm, 2.69 fm, and 2.43 fm, which are slightly larger than 
the experimental values,  2.39 fm and 2.22 fm, of $^9$Be and $^{10}$Be
reduced from charge radii.  
The calculated magnetic and electric quadrupole moments 
of $^9$Be are $\mu=-1.06$ $\mu_N$ and 
$Q=6.9$  e\ fm$^2$, which reasonably agree to the experimental values, 
 $\mu=-1.1778(9)$ $\mu_N$ and 
$Q=5.288(38)$  e\ fm$^2$.

\subsection{Excited states}
\begin{figure}[htb]
\begin{center}
\includegraphics[width=8cm]{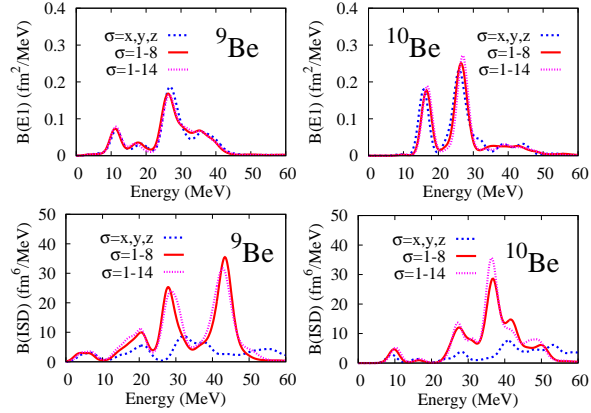} 	
\end{center}
  \caption{(color online) 
E1 and ISD strengths of $^9$Be and $^{10}$Be 
obtained by the sAMD in three cases of 
$\bvec{e}_{\sigma=x,y,z}$, $\bvec{e}_{\sigma=1,\ldots,8}$, and $\bvec{e}_{\sigma=1,\ldots,14}$.
The smearing width is $\gamma=2$ MeV.
\label{fig:be-sp14}}
\end{figure}

To investigate dipole excitations,  
I calculate $J^\pi=1^-$ states of $^8$Be and $^{10}$Be, and 
$J^\pi=1/2^+$, $3/2^+$, and $5/2^+$ states of $^9$Be
by applying the sAMD and the sAMD+$\alpha$GCM based 
on the obtained ground state wave functions.
For the sAMD, the shift parameter $\epsilon$ is taken to be $\epsilon=0.1$ fm which is small
enough to give the $\epsilon$-independent result. 
For unit vectors $\bvec{e}_\sigma$, 
I choose three sets, $\bvec{e}_{\sigma=x,y,z}$, $\bvec{e}_{\sigma=1,\ldots,8}$, and $\bvec{e}_{\sigma=1,\ldots,14}$, 
and check the convergence of the dipole strengths.
Here, the set of 8 vectors are  
$\bvec{e}_{\sigma=1,\ldots,8}=(\pm 1/\sqrt{3},\pm 1/\sqrt{3},\pm 1/\sqrt{3})$, 
and that of 14 vectors are 
$\bvec{e}_{\sigma=1,\ldots,14}=(\pm 1/\sqrt{3}, \pm 1/\sqrt{3}, \pm 1/\sqrt{3})$, 
$(\pm 1, 0, 0)$, $(0, \pm 1, 0)$, and $(0, 0, \pm 1)$.
The $x$, $y$, and $z$ axes are taken to be the principle axes of the inertia of the intrinsic state that 
satisfy $\langle x^2\rangle \ge \langle y^2\rangle\ge \langle z^2\rangle$ 
and $\langle xy\rangle=\langle yz\rangle=\langle zx\rangle=0$.
For the E1 strength, the sAMD model space with $\bvec{e}_{\sigma=x,y,z}$ can cover 
 $\bvec{r}_j\phi_{{\bvec{X}}_i}({\bvec{r}_j})$ configurations 
excited by the E1 operator. However, 
for the ISD strength, a larger number of unit vectors ($\bvec{e}_{\sigma}$) 
are necessary for the sAMD model space to cover $r_j^2\bvec{r}_j\phi_{{\bvec{X}}_i}({\bvec{r}_j})$ configurations 
excited by the ISD operator. 
The E1 and ISD strengths of $^{9}$Be and $^{10}$Be calculated by the sAMD 
in three cases, $\bvec{e}_{\sigma=x,y,z}$, $\bvec{e}_{\sigma=1,\ldots,8}$, and $\bvec{e}_{\sigma=1,\ldots,14}$,
are shown in Fig.~\ref{fig:be-sp14}. 
As expected, the set $\bvec{e}_{\sigma=x,y,z}$ is enough only for the E1 strength but 
not for the ISD strength.
It is found that the set $\bvec{e}_{\sigma=1,\ldots,8}$ is practically enough to get 
a qualitatively converged result for both the E1 and ISD strengths, and therefore
this set is adopted in the present calculation of the dipole strengths.
For the $\alpha$GCM calculation, 
the distance parameter  is taken to be $\Delta D=-1,1,2,\ldots,20$ fm. This means that
$\alpha$-cluster continuum states are treated as discretized states in the box boundary $\Delta D \le 20$ fm. 

The model space of the sAMD+$\alpha$GCM wave function given in Eq.~\eqref{eq:sAMD+aGCM}
covers 1p-1h excitations and  
$\alpha$-cluster excitations from the ground state wave function. 
For the detailed description of the
low-lying energy spectra and their dipole transition strengths, I mix additional configurations 
optimized for the low-lying levels 
$^9$Be$(1/2^+_1)$, $^9$Be$(3/2^+_1)$, $^9$Be$(5/2^+_1)$
and $^{10}$Be$(1^-_1)$, which are obtained by the AMD+VAP 
with $J^\pi=1/2^+$, $3/2^+$, and 
$5/2^+$ for $^9$Be and that with $J^\pi=1^-$ for $^{10}$Be. 
The final wave function with these additional VAP configurations is given as
\begin{eqnarray}
\Psi^{\textrm{sAMD+}\alpha\textrm{GCM+cfg}}_{{\rm Be}(J^\pi_k)}&=&
\sum_K c_0(J^\pi_k; K) 
P^{J\pi}_{MK}\Phi_{\rm AMD}({\bf Z}^0)\\
&&+
\sum_{i=1,\ldots,A}\sum_{\sigma}\sum_K 
c_1(J^\pi_k; i,\sigma,K) P^{J\pi}_{MK}\Phi_{\rm AMD}({\bf Z}_{\rm nonflip}^0(i,\sigma))\nonumber\\
&&+
\sum_{i=1,\ldots,A}\sum_{\sigma}\sum_K 
c_2(J^\pi_k; i,\sigma,K) P^{J\pi}_{MK}\Phi_{\rm AMD}({\bf Z}_{\rm flip}^0(i,\sigma))\nonumber\\
&&+
\sum_{\Delta D}\sum_K c_3(J^\pi_k; \Delta D,K) 
P^{J\pi}_{MK}\Phi_{\rm AMD}(\bvec{Z}^0_{D_\alpha}(\Delta D))\nonumber \\
&+&\sum_{J'\pi'}\sum_K c_4(J^\pi_k;J'\pi',K) 
P^{J\pi}_{MK}\Phi_{\rm AMD}(\bvec{Z}^{J'\pi'}_\textrm{VAP}).
\end{eqnarray}
The dipole strengths are calculated by the following matrix elements, 
\begin{eqnarray}
&&\langle\Psi^{\textrm{sAMD}}_{{\rm Be}(J^\pi_k)}||
{\cal M}||\Psi^{\textrm{sAMD}}_{{\rm Be}(\textrm{g.s.})}\rangle,\\
&&\langle\Psi^{\textrm{sAMD+}\alpha\textrm{GCM}}_{{\rm Be}(J^\pi_k)}||
{\cal M}||\Psi^{\textrm{sAMD+}\alpha\textrm{GCM}}_{{\rm Be}(\textrm{g.s.})}\rangle,\\
&&\langle\Psi^{\textrm{sAMD+}\alpha\textrm{GCM+cfg}}_{{\rm Be}(J^\pi_k)}||
{\cal M}||\Psi^{\textrm{sAMD+}\alpha\textrm{GCM+cfg}}_{{\rm Be}(\textrm{g.s.})}\rangle.
\end{eqnarray}
The calculated dipole strengths with and without the additional VAP configurations 
are found to be almost consistent with each other except for quantitative details of the energy position 
and the strengths in $E \le 10$ MeV.
In this paper, I mainly discuss the dipole strengths calculated by the sAMD
and the sAMD+$\alpha$GCM+cfg wave functions, which I call cal-I and cal-II, respectively.
The former corresponds to the small amplitude calculation containing 
1p-1h excitations.  The latter contains 
the large amplitude $\alpha$-cluster mode in addition to the 1p-1h excitations described by the sAMD model space.
Namely, the sAMD+$\alpha$GCM+cfg  includes the 
higher correlation than the sAMD 
in both the ground and excited states.

\subsection{E1 strength}

\begin{figure}[htb]
\begin{center}
\includegraphics[width=6.0cm]{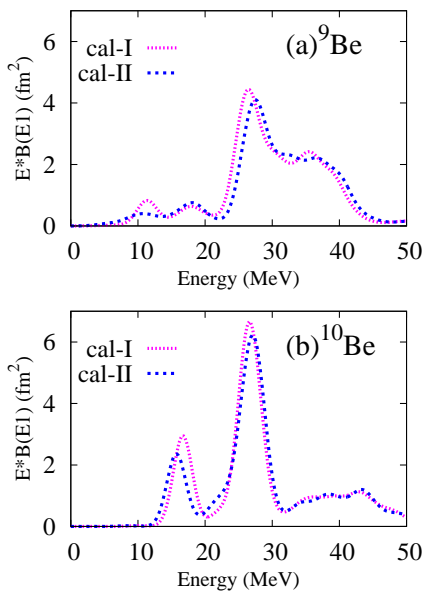} 	
\end{center}
  \caption{(color online) 
Energy-weighted E1 strength of $^9$Be and $^{10}$Be 
obtained by the sAMD (cal-I) and the sAMD+$\alpha$GCM+cfg (cal-II).
The smearing width is $\gamma=2$ MeV.
\label{fig:beiso-e1-gcm}}
\end{figure}

\begin{figure}[htb]
\begin{center}
\includegraphics[width=6.0cm]{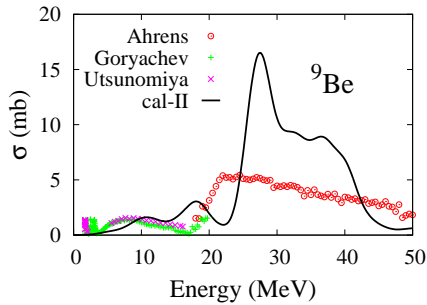} 	
\end{center}
   \caption{(color online) Comparison of the calculated E1 cross section of $^9$Be 
with the experimental photonuclear cross sections.
The calculated values are those obtained with 
the sAMD+$\alpha$GCM+cfg (cal-II), smeared by $\gamma=2$ MeV.
The experimental data are taken from the 
photonuclear cross sections by Ahrens {\it et al.}\cite{Ahrens:1975rq}, the 
bremsstrahlung data by Goryachev {\it et al.}\cite{Goryachev92}, and the 
photodisintegration cross sections by Utsunomiya {\it et al.}\cite{Utsunomiya15}.
\label{fig:be9-exp}}
\end{figure}

The energy-weighted E1 strength of $^9$Be and $^{10}$Be 
obtained by the sAMD (cal-I) and the sAMD+$\alpha$GCM+cfg (cal-II)
is shown in Fig.~\ref{fig:beiso-e1-gcm}.
The strength functions of two calculations (I) and (II) are qualitatively similar to each other
except for broadening of the low-energy strength in $E\le 15$ MeV of $^9$Be in
the cal-II.
The calculated EWS of the E1 strength is enhanced from the TRK sum rule value
by a factor $1.7-1.8$.
Figure \ref{fig:be9-exp} shows the comparison of the calculated E1 cross section 
with the experimental photonuclear cross sections of $^9$Be. 
The calculation reasonably describes the global feature of the
experimental cross sections consisting of    
the low-energy strength in $E<20$ MeV and the GDR in $E>20$ MeV, 
though it somewhat overestimates the GDR peak energy and strength.

\begin{figure}[htb]
\begin{center}
\includegraphics[width=6.0cm]{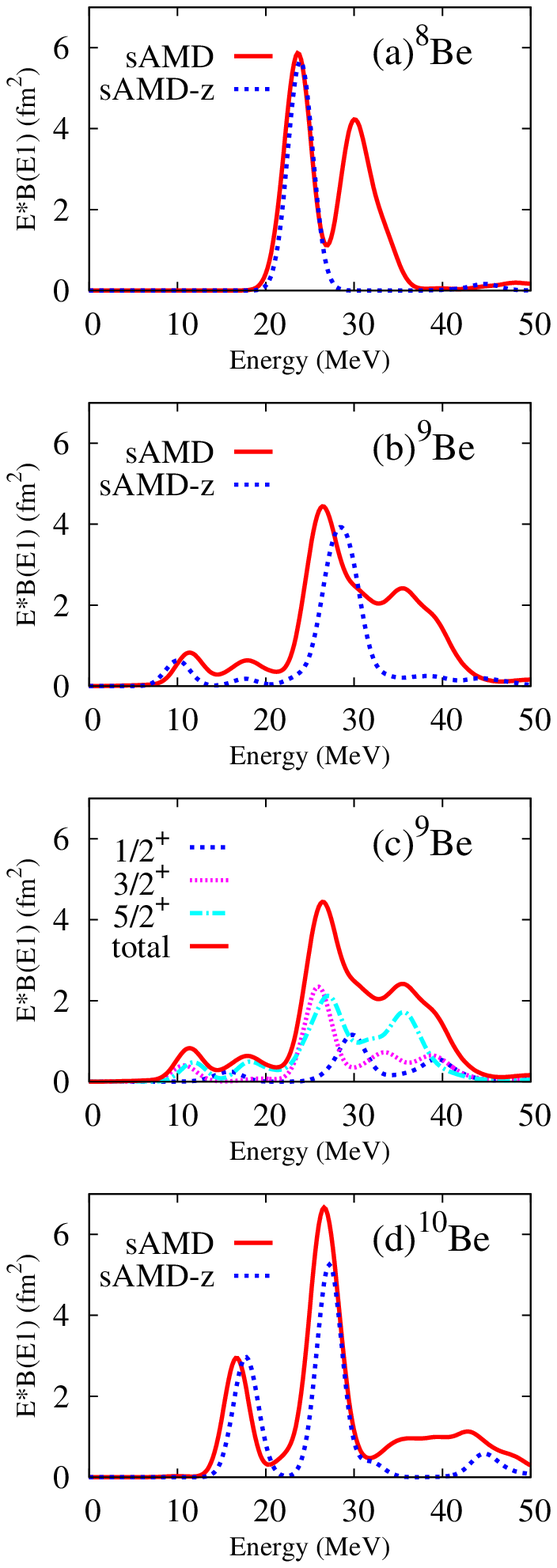} 	
\end{center}
  \caption{(color online)  
Energy-weighted E1 strength of $^8$Be, $^9$Be, and $^{10}$Be obtained by the 
sAMD ($\sigma=1,\ldots,8$) and the sAMD-z (longitudinal mode: $\sigma=z$) calculations. 
The sAMD and sAMD-z results for (a)$^8$Be, (b)$^9$Be, and (d)$^{10}$Be
are shown by solid and dashed lines, respectively. 
The decomposition of the strengths for $J^\pi=1/2^+$, $3/2^+$, and $5/2^+$ states of 
$^9$Be is also shown as well as the total strength in the panel (c).
The smearing width is $\gamma=2$ MeV. 
\label{fig:beiso-e1}}
\end{figure}

As shown in the comparison of the sAMD (cal-I) and the sAMD+$\alpha$GCM+cfg (cal-II)
in Fig.~\ref{fig:beiso-e1-gcm}, the E1 strength of $^9$Be and $^{10}$Be is 
not affected so much by the coupling with the large amplitude $\alpha$-cluster motion.
In the following, we give detailed analysis of the E1 strength of $^8$Be, $^9$Be, and $^{10}$Be 
based on the sAMD to discuss effects of excess neutrons on the E1 strength in $^9$Be and $^{10}$Be.
Figure \ref{fig:beiso-e1} shows the E1 strength
of $^8$Be, $^9$Be, and $^{10}$Be calculated by the sAMD.
For $^9$Be, decomposition of the transition strength to $J=1/2^+$, $3/2^+$, and $5/2^+$ states is also shown.
The GDR in $^8$Be shows a two peak structure in $E=20-40$ MeV. Also in $^9$Be,
the two peak structure of the GDR is seen but it somewhat broadens.
In addition to the GDR, low-lying E1 strength appears in $E=10-20$ MeV of $^9$Be.
In $^{10}$Be, the lower peak of the GDR exists at $E\sim 25$ MeV, whereas 
the higher peak of the GDR is largely fragmented. Below the GDR, 
an E1 resonance appears at $E\sim 15$ MeV. 

The origin of the two-peak structure of the GDR  in Be isotopes is the
prolate deformation of the $2\alpha$ core. To distinguish the longitudinal mode in the intrinsic frame, 
we calculate the E1 strength in the truncated sAMD model space by using wave functions shifted only 
to the longitudinal ($z$) direction, that is, the sAMD with the fixed $\sigma=z$ as
\begin{eqnarray}
\Psi^{\textrm{sAMD-z}}_{{\rm Be}(J^\pi_k)}&=& c_0(J^\pi_k) 
P^{J\pi}_{MK}\Phi_{\rm AMD}({\bf Z}^0)\\
&&+\sum_{i=1,\ldots,A}
(c_1(J^\pi_k; i) P^{J\pi}_{M0}\Phi_{\rm AMD}({\bf Z}_{\rm nonflip}^0(i,z))
+c_2(J^\pi_k; i) P^{J\pi}_{M0}\Phi_{\rm AMD}({\bf Z}_{\rm flip}^0(i,z)),
\end{eqnarray}
where the coefficients $c_0$, $c_1$, and $c_2$ are determined by diagonalization 
of the norm and Hamiltonian matrices. Here I omit the $K$-mixing and fix $K=0$ for $^{8}$Be and $^{10}$Be, 
and $K=3/2$ for $^{9}$Be to take into account only the $Y^1_0$ mode in the intrinsic frame. 
The sAMD with $\sigma=z$, which I call ``sAMD-z'',  is approximately regarded as 
the calculation containing the longitudinal mode but no transverse mode, though 
two modes do not exactly decouple from each other because of the angular-momentum projection.
The E1 strength obtained by the sAMD-z is shown by dashed lines
in Fig.~\ref{fig:beiso-e1}(a), (b), and (d). In comparison of the sAMD and  sAMD-z results,  
it is found that the lower peak of the GDR at $E=20-30$ MeV  
is contributed by the longitudinal mode of the $2\alpha$ core, whereas the higher peak of the GDR
comes from the transverse mode.  
The higher peak broadens in $^9$Be and it is largely fragmented in $^{10}$Be 
indicating that the transverse mode is affected by
excess neutrons. For the low-lying E1 resonances below the GDR, 
the strength at $E \sim 10$ MeV in $^{9}$Be and that 
at $E \sim 15$ MeV  in $^{10}$Be are mainly contributed by the 
longitudinal mode. These low-energy dipole resonances in $^{9}$Be and  $^{10}$Be
are understood by the longitudinal motion of valence neutrons against the $2\alpha$ core. 

From the above analysis of 
the E1 strength of $^9$Be and $^{10}$Be compared with that of $^8$Be, 
the effects of excess neutrons on the E1 strength is understood as follows.
The longitudinal and transverse dipole modes in the $2\alpha$ core part contribute
to the GDR with the two peak structure. 
The valence neutron modes couple with the transverse dipole mode
of the $2\alpha$ core and they broaden the higher peak of the GDR. 
Moreover, the  valence neutron modes against the $2\alpha$ core 
contribute to the low-energy E1 strength.
More details of the low-energy dipole excitations are discussed later. 

\subsection{ISD strength and coupling with the $\alpha$-cluster mode in $^9$Be and $^{10}$Be}
\begin{figure}[htb]
\begin{center}
\includegraphics[width=6.0cm]{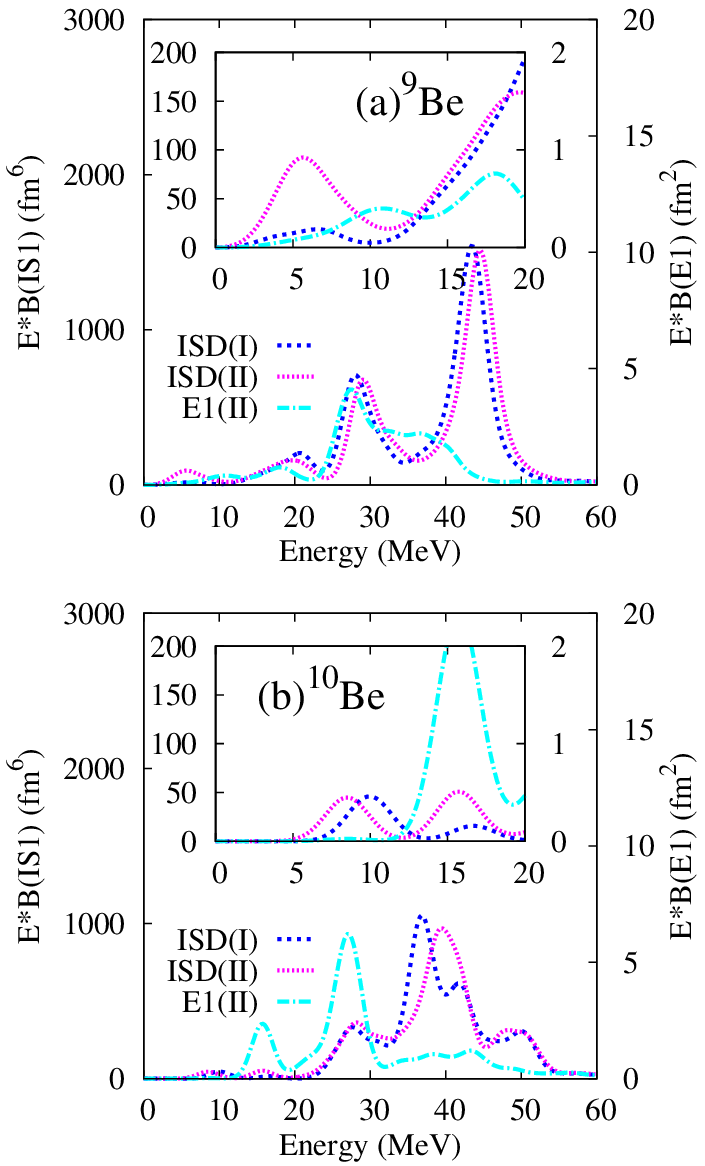} 	
\end{center}
  \caption{(color online) 
ISD strength of $^9$Be and $^{10}$Be obtained by the sAMD (cal-I) and 
the sAMD+$\alpha$GCM+cfg (cal-II). E1 strength obtained by the 
the sAMD+$\alpha$GCM+cfg (cal-II) is also shown for comparison.
The smearing width is $\gamma=2$ MeV.
\label{fig:be-isd}}
\end{figure}

As previously mentioned, the sAMD (cal-I) corresponds to 
the small amplitude calculation, whereas the sAMD+$\alpha$GCM+cfg (cal-II) contains
the large amplitude $\alpha$-cluster mode.
A possible enhancement of the ISD strength in
the cal-II relative to the cal-I can be a good probe
for the dipole excitation that couples with the $\alpha$-cluster mode,
because the $\alpha$-cluster excitation in $^9$Be and $^{10}$Be involves 
the compressive dipole mode. 
The energy-weighted ISD strength of $^9$Be and $^{10}$Be calculated by the 
cal-I and cal-II
is shown in Fig.~\ref{fig:be-isd}.
The strength of the isoscalar GDR in $E=30\sim 50$ MeV
is not affected by the $\alpha$-cluster mode, 
whereas the ISD strength for some low-energy resonances are significantly enhanced
in the cal-II as a result of the coupling with the $\alpha$-cluster mode.
In $^9$Be, the ISD strength in $E<10$ MeV is remarkably enhanced 
in the cal-II, whereas  
the resonance at $E=10-15$ MeV has the weak ISD strength in both the cal-I and cal-II.
In $^{10}$Be, the ISD strength around $E=15$ MeV is enhanced. 

\subsection{Low-energy dipole resonances in $^9$Be and $^{10}$Be}

\begin{table}[htb]
\caption{EWS of the E1 and ISD strengths for low-energy 
resonances: A1($E<8$ MeV), A2($8<E<15$ MeV), and A3($15<E<20$ MeV) in $^9$Be,
and B1($E<12$ MeV) and B2($12<E<20$ MeV) in $^{10}$Be. The total EWS is also shown.
The strengths are calculated by the sAMD (cal-I) and 
the sAMD+$\alpha$GCM+cfg (cal-II). The EWS calculated by the matrix elements 
for 
the transitions from the sAMD+$\alpha$GCM+cfg  initial states to 
the sAMD final states (cal-III), and  that for the transitions 
from the sAMD initial states to sAMD+$\alpha$GCM+cfg 
the final states (cal-IV)  are also listed. The unit of $S(E1)$ is fm$^2$MeV and that of  $S(ISD)$ is fm$^6$MeV.
\label{tab:EWS} 
 }
\begin{center}
\begin{tabular}{ccccc}
\hline
  &  (I)  & (II) & (III) & (IV) \\
$\alpha$ mode in initial & w/o & w & w & w/o\\
$\alpha$ mode in final & w/o & w & w/o & w \\
 \multicolumn{5}{c} {$^9$Be} \\
$S(E1;\textrm{total})$	&	58 	&	56 	&	58 	&	56 	\\
$S(E1;\textrm{A1})$	&	0.13	&	0.43	&	0.39	&	0.16	\\
$S(E1;\textrm{A2})$	&	3.3 	&	2.4 	&	2.6 	&	3.5 	\\
$S(E1;\textrm{A3})$	&	2.7 	&	3.3 	&	3.3 	&	2.6 	\\
$S(IS1;\textrm{total})$	&	$15.2\times 10^3$ 	&	$15.3\times 10^3$  	&	$15.2\times 10^3$ 	& $15.3\times 10^3$ 	\\
$S(IS1;\textrm{A1})$	&	93 	&	410 	&	58 	&	124 	\\
$S(IS1;\textrm{A2})$	&	108 	&	230 	&	200 	&	131 	\\
$S(IS1,\textrm{A3})$	&	490 	&	560 	&	640 	&	520 	\\									
 \multicolumn{5}{c}{$^{10}$Be} \\															
$S(E1;\textrm{total})$	&	63 	&	62 	&	63 	&	62 	\\
$S(E1;\textrm{B1})$	&	0.09	&	0.08	&	0.06	&	0.10	\\
$S(E1;\textrm{B2})$	&	10.4 	&	8.7 	&	7.6 	&	10.9 	\\
$S(IS1;\textrm{total})$	&	$12.7\times 10^3$ 	&	$13.0\times 10^3$ 	&	$12.7\times 10^3$ 	&	$13.0\times 10^3$	\\
$S(IS1;\textrm{B1})$	&	162 	&	157 	&	115 	&	187	\\
$S(IS1;\textrm{B2})$	&	56 	&	187 	&	83 	&	74 	\\
\hline	
\end{tabular}
\end{center}
\end{table}

From the analysis of the E1 and ISD strengths, the low-energy dipole excitations below the GDR
in $^9$Be can be 
categorized as three resonances in $E<8$ MeV, $8<E<15$ MeV, and 
$15<E<20$ MeV, which I label A1, A2, and A3 resonances, respectively. 
The EWS in the corresponding energy regions, $S(E1/IS1;\textrm{A1})$, $S(E1/IS1;\textrm{A2})$, $S(E1/IS1;\textrm{A3})$,  is listed in Table \ref{tab:EWS}
as well as the total EWS value.
In addition to the EWS obtained by the sAMD (cal-I) and 
sAMD+$\alpha$GCM+cfg (cal-II), 
the EWS calculated by the matrix elements
\begin{equation}
\langle\Psi^{\textrm{sAMD}}_{{\rm Be}(J^\pi_k)}|{\cal M}|
\Psi^{\textrm{sAMD+}\alpha\textrm{GCM+cfg}}_{{\rm Be}(\textrm{g.s.})}\rangle
\end{equation}
for the transitions from the sAMD+$\alpha$GCM+cfg initial states to 
the sAMD final states (cal-III) and that by the matrix elements
\begin{equation}
\langle\Psi^{\textrm{sAMD+}\alpha\textrm{GCM+cfg}}_{{\rm Be}(J^\pi_k)}|{\cal M}|
\Psi^{\textrm{sAMD}}_{{\rm Be}(\textrm{g.s.})}\rangle
\end{equation}
for the transitions from the sAMD initial states to the sAMD+$\alpha$GCM+cfg 
final states (cal-IV) are also 
shown in the table. The cal-III contains the $\alpha$-cluster mode
in the initial states as the ground state correlation 
but not in the final states, whereas the cal-IV contains the 
$\alpha$-cluster mode only in the final states but not in the initial states.
In all the calculations (I), (II), (III), and (IV), $S(E1;\textrm{A1})$
is very small, whereas 
$S(E1;\textrm{A2})$ and $S(E1;\textrm{A3})$
are significantly large as $\sim$10\% of the TRK sum rule. 
Consequently, the EWS of the E1 strength in $E<20$ MeV 
exhausts $\sim$20\% of the TRK sum rule and it is $\sim$10\% of the calculated total EWS.

The $\alpha$-cluster mode does not affect so much the E1 and ISD strengths of
A2 and those of A3, but it gives significant enhancement of the dipole strengths of
A1. In particular,  $S(IS1;\textrm{A1})$  is remarkably enhanced 
by the coupling with the $\alpha$-cluster mode. The
enhancement is found only in the cal-II but not in other calculations, cal-I,  cal-III, and cal-IV. 
It indicates that the coupling with the $\alpha$-cluster mode in the ground state and that in the A1 resonance 
coherently enhance $S(IS1;\textrm{A1})$. 
The $\alpha$-cluster mode also makes  $S(E1;\textrm{A1})$
 three times larger in the cal-II
than the cal-I, though it is still less than 2\% of the TRK sum rule.

\begin{figure}[htb]
\begin{center}
\includegraphics[width=7.5cm]{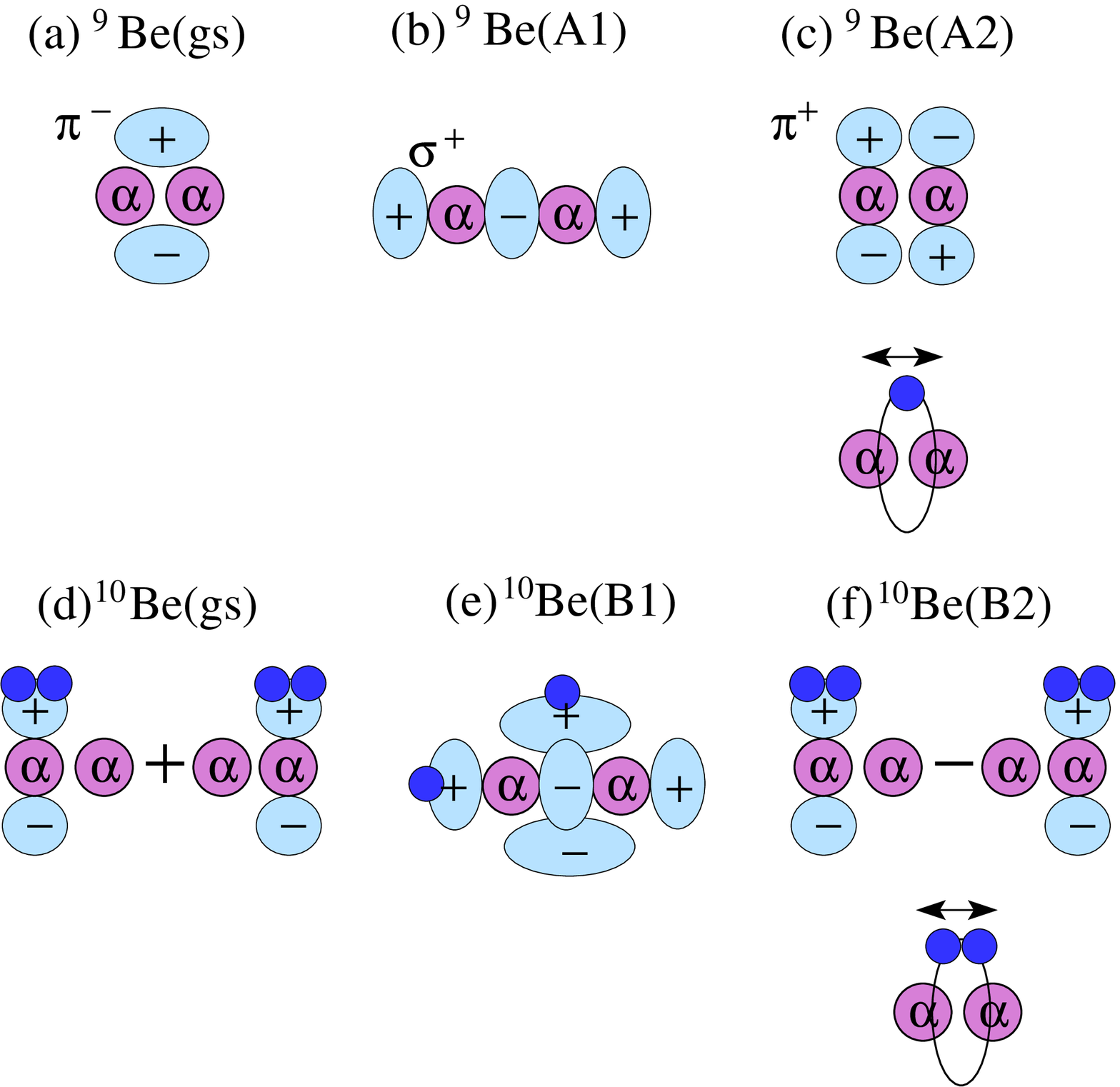} 	
\end{center}
  \caption{(color online)  Schematic figures of $2\alpha$ and valence neutrons for
the ground state and the A1 and A2 resonances of $^9$Be,
and those for the ground state and the B1 and B2 resonances of $^{10}$Be. 
\label{fig:orbital}}
\end{figure}

As seen in the EWS of the sAMD+$\alpha$GCM+cfg (cal-II) in Table \ref{tab:EWS}, 
the A1 resonance shows the relatively strong ISD and weak E1 transitions, 
whereas the A2 resonance shows the relatively weak ISD and strong E1 transitions. 
These characteristics of the A1 and A2 resonances 
can be understood by the $2\alpha+n$ picture
as follows.
The ground and low-lying states of $^9$Be are approximately described by 
the molecular orbital structure, where the valence neutron occupies molecular orbitals formed 
by the linear combination of 
$p$ orbits around $\alpha$ clusters 
\cite{Okabe77,Okabe78,SEYA,OERTZENa}.
Let me consider two $\alpha$ clusters at the left and right along
the $z$ axis (see Fig.~\ref{fig:orbital}). I call the left(right) $\alpha$ ``$\alpha_{L(R)}$'', 
and label single-particle orbits (atomic orbitals) around each $\alpha$ cluster 
as $[Nn_z l_z j_z]_{\alpha_{L(R)}}$. 
Here $N$ is the total quantum (node) number, 
$n_z$ is the quantum number for the $z$-axis, and 
$l_z$ and $j_z$ are the $z$-components of the orbital- and total-angular momenta, respectively.
The $\pi^-_{3/2}$ and $\pi^+_{3/2}$ molecular orbitals are given by the linear combination of the 
atomic orbitals
$[101\frac{3}{2}]_{\alpha_{L,R}}$ as
\begin{eqnarray}
&&\pi^-_{3/2}\equiv[101\frac{3}{2}]_\textrm{MO}=[101\frac{3}{2}]_{\alpha_{L}}+ [101\frac{3}{2}]_{\alpha_{R}}\\
&&\pi^+_{3/2}\equiv[211\frac{3}{2}]_\textrm{MO}=[101\frac{3}{2}]_{\alpha_{L}}- [101\frac{3}{2}]_{\alpha_{R}}\\
\end{eqnarray}
where $[Nn_zl_zj_z]_\textrm{MO}$ is the label indicating the quantum numbers $N$, $n_z$, $l_z$, and $j_z$ of the 
molecular orbital around the $2\alpha$.
Another molecular orbital is the longitudinal orbital $\sigma^+_{1/2}$ given by the linear combination of
 $[110\frac{1}{2}]_{\alpha_{L,R}}$ as
\begin{eqnarray}
&&\sigma^+_{1/2}\equiv[220\frac{3}{2}]_\textrm{MO}
=[110\frac{1}{2}]_{\alpha_{L}}-[110\frac{1}{2}]_{\alpha_{R}}.
\end{eqnarray}
In the case that the $\alpha$-$\alpha$ distance is not so large, the molecular orbital $[Nn_zl_zj_z]_\textrm{MO}$
approximately corresponds to the Nilsson (deformed shell-model) orbit $[Nn_3\Lambda]_\Omega$ with 
$n_3=n_z$, $\Lambda=l_z$, and $\Omega=j_z$ in the prolate deformation. 
$\pi^-_{3/2}$ is the negative-parity orbital with no node ($n_z=0$), 
$\pi^+_{3/2}$ is the positive-parity orbital with one node ($n_z=1$), 
and $\sigma^+_{1/2}$ is the positive-parity orbital with two nodes ($n_z=2$) along the $z$-axis.

For the valence neutron around the $2\alpha$,  $\pi^-_{3/2}$ is the lowest negative-parity molecular orbital,
whereas $\sigma^+_{1/2}$ is the lowest positive-parity orbital.
The ground state of $^9$Be dominantly has the $2\alpha+n$ structure with the  $\pi^-_{3/2}$
configuration. 
The A1 resonance is approximately described by the $\sigma^+_{1/2}$ configuration, whereas
the A2 resonance  is dominated by the  $\pi^+_{3/2}$ configuration
(see Fig.~\ref{fig:orbital}). 
The E1 transition for $\pi^-_{3/2}\to \pi^+_{3/2}$, i.e., 
$[101\frac{3}{2}]_\textrm{MO}\to [211\frac{3}{2}]_\textrm{MO}$ 
is possible because the $Y^1_0$ operator changes $N\to N\pm 1$ and $n_z\to n_z\pm 1$.  However, 
the E1 transition for $\pi^-_{3/2}\to \sigma^+_{1/2}$, i.e., $[101\frac{3}{2}]_\textrm{MO}\to [220\frac{1}{2}]_\textrm{MO}$
is forbidden because the change $n_z\to n_z\pm 2$ is not possible for the E1 operator.
This is the reason why the E1 strength is large for A2 but it is suppressed for A1.
Because of the $[211\frac{3}{2}]_\textrm{MO}$ configuration of the A2 resonance, 
the E1 strength of A2 shows the $K=3/2$ band feature that the contribution from
transitions to $J^\pi=3/2^+$ and $5/2^+$ states is dominant 
as seen in Fig.~\ref{fig:beiso-e1}(c).
The A1 resonance has the large node number $n_z=2$ along the $2\alpha$ direction than 
the A2 resonance ($n_z=1$), and therefore, the spatial development of the $2\alpha$ clustering is 
more prominent in the A1 resonance. 
As a result of the developed clustering, the A1 resonance couples rather strongly with the $\alpha$-cluster mode. 
The coupling with the $\alpha$-cluster mode, namely, the $^5$He-$\alpha$ relative motion in A1
enhances the ISD strength as discussed previously.

Let me discuss low-energy dipole resonances in $^{10}$Be. 
The low-energy dipole strength below the GDR can be categorized as two resonances in $E<12$ MeV and 
$12 < E < 20$ MeV, which I label B1 and B2, respectively. The EWS of the dipole strengths 
for the corresponding energy regions are listed in table \ref{tab:EWS}. 
The B2 resonance shows the strong E1 transition exhausting more than 20\% of the TRK sum rule and 10\% of 
the calculated total EWS. It also shows the significant ISD strength 
enhanced by the $\alpha$-cluster mode in the cal-II (sAMD+$\alpha$GCM+cfg).
The significant E1 strength and 
the strong coupling with the $\alpha$-cluster mode of the B2 resonance can be understood by 
two neutron correlation in the $2\alpha+2n$ picture as shown in the schematic figures of Fig.~\ref{fig:orbital}. 
The configuration in the ground state of $^{10}$Be is approximately described by the positive-parity 
projected state of the atomic orbital configuration  as
\begin{eqnarray}
&&[101\frac{3}{2}]_{\alpha_{L}}[10-1-\frac{3}{2}]_{\alpha_{L}}+ [101\frac{3}{2}]_{\alpha_{R}}
[10-1-\frac{3}{2}]_{\alpha_{R}}\\
&&=\frac{1}{2}\left\{  (\pi^-_{3/2})^2+(\pi^+_{3/2})^2) \right\},
\end{eqnarray}
which corresponds to the $^6$He+$\alpha$ cluster structure in the intrinsic state of the $^{10}$Be ground state.
The B2 resonance is interpreted as the parity partner of the ground state as 
\begin{eqnarray}
&&[101\frac{3}{2}]_{\alpha_{L}}[10-1-\frac{3}{2}]_{\alpha_{L}} - [101\frac{3}{2}]_{\alpha_{R}}
[10-1-\frac{3}{2}]_{\alpha_{R}}\\
&&=\frac{1}{2}\left\{ \pi^-_{3/2}\pi^+_{3/2} + \pi^+_{3/2}\pi^-_{3/2}\right\}.
\end{eqnarray}
In the transition from the ground state to the B2 resonance, the coherent contribution of two neutrons 
enhances the E1 strength.
The B2 resonance has a large overlap with negative-parity $^6$He+$\alpha$ cluster wave functions, 
for instance, 60\% overlap with $P^{1-}_{00}\Phi_{\rm AMD}({\bf Z}^0_{D_\alpha}(\Delta D=1 \textrm{fm}))$. 
As a result of the strongly coupling with the $\alpha$-cluster mode, the ISD strength of the B2 resonance is enhanced. 
In other words, the B2 resonance is regarded as the $\alpha$-cluster excitation on the ground state which already
contains the $^6$He+$\alpha$ cluster structure.
In contrast to the B2 resonance,
the B1 resonance is regarded as a single-particle excitation with the molecular orbital configuration
$\pi^-_{3/2}\sigma^+_{1/2}$ and has no coherent contribution of two neutrons to the E1 strength.
Moreover, the $\pi^-_{3/2}\sigma^+_{1/2}$ configuration 
contains the $^5$He+$^5$He and $^6$He$^*$+$\alpha$ 
components instead of the $^6$He+$\alpha$ component, and therefore, it does not couple with 
the $\alpha$-cluster mode. 

\begin{figure}[htb]
\begin{center}
\includegraphics[width=6.0cm]{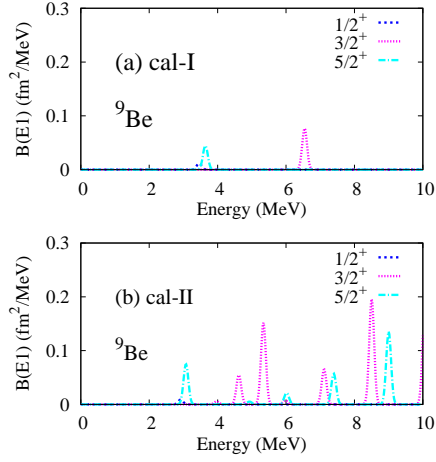} 	
\end{center}
  \caption{(color online)  E1 strength of $1/2^+$,  $3/2^+$,  and $5/2^+$ states of
$^9$Be in $E< 10$ MeV obtained by (a) the sAMD (cal-I) and (b)
the sAMD+$\alpha$GCM+cfg (cal-II). 
The smearing width is $\gamma=0.1$ MeV.
\label{fig:be9-e1}}
\end{figure}

Finally, I discuss the calculated $B(E1)$ of the $1/2^+_1$,  $3/2^+_1$,  and $5/2^+_1$ states of 
$^9$Be, which contribute to the dipole strengths of the A1 resonance. 
In Fig.~\ref{fig:be9-e1}, I show the E1 strength in $E\le 10$ MeV of $^9$Be. Here the smearing width is chosen to be $\gamma=0.1$ MeV to resolve discrete states.
The $1/2^+_1$,  $3/2^+_1$,  and $5/2^+_1$ states are obtained as discrete states in the sAMD (cal-I).
In the sAMD+$\alpha$GCM+cfg (cal-II), 
the $1/2^+_1$ and $5/2^+_1$ states are still discrete states, however, the 
$3/2^+_1$ shows a resonance behavior coupling with the discretized continuum states
in the box boundary at $\Delta D\le 20$ fm.  
We evaluate the $B(E1)$ of the $3/2^+_1$ resonance by a sum of the E1 strength in $E< 6$ MeV 
and estimate the excitation energy by the $B(E1)$ weighted 
averaged energy of  $3/2^+$ states in this energy region. 
In Table \ref{tab:BE1}, I compare the calculated E1 strength of the $1/2^+_1$,  $3/2^+_1$,  and $5/2^+_1$ states with the experimental values measured by the  $(\gamma,n)$ cross sections in Ref.~\cite{Arnold:2011nv}.
The excitation energies and $B(E1)$ of  the $^9$Be($5/2^+_1$)  and  $^9$Be($3/2^+_1$) obtained 
by the sAMD+$\alpha$GCM+cfg (cal-II) reproduce reasonably the experimental data. 
However, the calculated $B(E1)$ 
of the $^9$Be($1/2^+_1$) is quite small inconsistently to the experimental data.
The $1/2^+_1$ state has been suggested to be a virtual or resonance state of a $s$-wave neutron
\cite{Efros:1998yk,Arai:2003jm,Burda:2010an,Garrido:2010xz,AlvarezRodriguez:2010ng,Efros:2013zoa,Odsuren:2015pja}.
However, the present model is insufficient to describe such the virtual or $s$-wave resonance state 
because the valence neutron motion far from the $2\alpha$
is not taken into account in the model space. 

 \begin{table}[htb]
\caption{
\label{tab:BE1} 
Excitation energy and E1 strength of $^9$Be($1/2^+_1$),  $^9$Be($3/2^+_1$),  $^9$Be($5/2^+_1$),
and $^{10}$Be($1^-_1$).  
The energy (MeV) and $B(E1)$ (fm$^2$) calculated by the sAMD (cal-I) and 
the sAMD+$\alpha$GCM+cgf (cal-II) are listed
compared with the experimental data.
The experimental values of $^9$Be are data measured by the $(\gamma,n)$ cross sections in
Ref.~\cite{Arnold:2011nv}. The experimental excitation energy of $^{10}$Be($1^-$) 
is taken from Refs.~\cite{AjzenbergSelove:1988ec,Tilley:2004zz}.
}
\begin{center}
\begin{tabular}{ccccccc}
\hline

  & \multicolumn{2}{c} { cal-I } & \multicolumn{2}{c}{ cal-II } &
\multicolumn{2}{c} {exp} \\
	&$E$	&	$B(E1)$	&	$E$&		$B(E1)$	&	$E$	&	$B(E1)$\\	
 $^9$Be & & & & & &  \\
$1/2^+$	&	3.5 	&	0.002 	&	2.9 	&	0.002 	&	1.731(2)	&	0.136(2)	\\
$3/2^+$	&	6.5 	&	0.014 	&	5.1 	&	0.039 	&	4.704	&	0.068(7)	\\
$5/2^+$	&	3.6 	&	0.008 	&	3.1 	&	0.013 	&	3.008(4)	&	0.016(2)	\\
 $^{10}$Be & & & & & &  \\
$1^-$	&	9.8 	&	0.009 	&	8.3 	&	0.010 	&	5.960 & \\
\hline	
\end{tabular}
\end{center}
\end{table}

\section{Summary} \label{sec:summary}
I investigated the isovector and isoscalar dipole excitations in $^9$Be and $^{10}$Be
with the shifted AMD combined with the $\alpha$-cluster GCM, in which 
the 1p-1h excitations on the ground state and the large amplitude $\alpha$-cluster mode are
incorporated. Since the angular-momentum and parity projections are done,  the
coupling of excitations in the intrinsic frame with the rotation and parity transformation is 
taken into account microscopically. 
The low-energy E1 resonances 
appear in $E<20$ MeV because of valence neutron modes against the $2\alpha$ core.
They exhaust about $20\%$ of the TRK sum rule and $10\%$ of the calculated EWS.
The GDR shows the two peak structure
which is understood by the E1 excitations in the 2$\alpha$ core part with the prolate deformation.
The higher peak of the GDR for the transverse mode 
broadens in $^9$Be and it is largely fragmented in $^{10}$Be because of excess neutrons.

By comparing the results of the shifted AMD combined with and without the $\alpha$-cluster GCM,
I investigated how the E1 and ISD 
strengths in $^9$Be and $^{10}$Be are affected by the large amplitude $\alpha$-cluster mode. 
The ISD strength is a good probe to identify the dipole resonances 
that couples with the $\alpha$-cluster mode because 
the $\alpha$-cluster mode in $^9$Be and $^{10}$Be involves the compressive dipole mode.
It was found that the ISD strength
for some low-energy resonances in $^9$Be and $^{10}$Be are
enhanced by the coupling with the $\alpha$-cluster mode,
whereas the E1 strength is not so sensitive to the coupling with the $\alpha$-cluster mode.
In $^9$Be, the ISD strength of the low-energy resonance in $E<10$ MeV is remarkable. 
In $^{10}$Be, the ISD strength at $E\sim 15$ MeV 
is enhanced by the coupling with the $\alpha$-cluster mode.
This resonance at $E\sim 15$ MeV in $^{10}$Be is regarded 
as the $\alpha$-cluster excitation on the ground state having the $^6$He+$\alpha$ structure 
and can be interpreted as the parity partner of the ground state.
The E1 transition of this resonance is also strong
because of the coherent contribution of two valence neutrons.

The calculated E1 strength of $^9$Be reasonably describes the global feature of
experimental photonuclear cross sections consisting of    
the low-energy strength in $E<20$ MeV and the GDR in $E>20$ MeV,
though it somewhat overestimates the GDR peak energy and its strength. 
For the low-lying positive-parity states of $^9$Be, 
the calculated excitation energies and $B(E1)$ of  the $^9$Be($5/2^+_1$)  and  $^9$Be($3/2^+_1$) 
reasonably agree to the experimental data.
However, the calculation fails to reproduce the experimental $B(E1)$ of 
the $^9$Be($1/2^+_1$) because the present model is insufficient to describe 
the detailed asymptotic behavior of the $s$-wave neutron in the $1/2^+_1$ state.

\section*{Acknowledgments} 
The author would like to thank Dr.~Utsunomiya and Dr.~Kikuchi for fruitful discussions.
The computational calculations of this work were performed by using the
supercomputer in the Yukawa Institute for theoretical physics, Kyoto University. This work was supported by 
JSPS KAKENHI Grant Number 26400270.

\end{document}